\documentclass[twocolumn,prd,nofootinbib,superscriptaddress,eqsecnum,tightenlines,8pt]{revtex4}

\usepackage{fancyhdr}
\usepackage{amsmath,amsfonts,amssymb}

\usepackage{color}
\definecolor{red}{rgb}{1,0,0}
\definecolor{gre}{rgb}{0,0.6,0}
\definecolor{blu}{rgb}{0,0,1}

\usepackage{mathrsfs}

\usepackage{ulem}

\newcommand{\be}{\begin{equation}}
\newcommand{\ee}{\end{equation}}
\newcommand{\ba}{\begin{eqnarray}}
\newcommand{\ea}{\end{eqnarray}}

\begin{document}

\title{Testing the Everett Interpretation of Quantum Mechanics with Cosmology}

\date{\today}

\author{Aur\'elien Barrau}
\email{Aurelien.Barrau@cern.ch}
\affiliation{
Laboratoire de Physique Subatomique et de Cosmologie, Universit\'e Grenoble-Alpes, CNRS-IN2P3\\
53,avenue des Martyrs, 38026 Grenoble cedex, France\\
}%



\begin{abstract}
In this brief note, we argue that contrarily to what is still often stated, the Everett many-worlds interpretation of quantum mechanics is not in principle impossible to test. It is actually not more difficult (but not easier either) to test than most other kinds of multiverse theories. We also remind why multiverse scenarios can be falsified.
\end{abstract}

\maketitle


\section{The Everett many world interpretation}

This brief article does not contain any fundamentally new material. It only aims at contradicting a naive --and in our view incorrect-- belief that seems to be dominant in the physics community: the unfalsifiability of the Everett "many-worlds" interpretation of quantum mechanics. Although the Everett interpretation is taken more and more seriously, the reason for this choice seems to be mostly aesthetic as most articles about the many-worlds view mention its fundamental untestability.\\

At the intuitive level, the Everett interpretation of quantum mechanics (see \cite{everett,dewitt} for historical articles) states that all possible alternate histories of a quantum system are actually real. Each of them takes place in a different universe, there is no collapse of the wave function. All that could have happened in our past but didn't happen here, did indeed happen in another world. All that can happen in the future will happen in different universes. Every possible quantum outcome is real in one universe. How strange it might look, this vision is now considered as one of the mainstream interpretation of quantum mechanics. Its obvious advantage is somehow to take quantum mechanics "literally" without adding by hands a non-unitary evolution mysteriously triggered by the measurement process.\\

Slightly more rigorously, the Everett interpretation could be said to rely on two hypothesis. Firstly that the wavefunction has an observer-independent objective existence and actually is the fundamental object. For a non-relativistic N-particle system the wavefunction is a complex-valued field in a 3N dimensional space. Secondly that the wavefunction obeys the standard linear deterministic wave equations at all times. The observer plays no special role in the theory and, consequently, there is no collapse of the wavefunction. For non-relativistic systems, the Schr\"odinger wave equation describes the evolution. \\

\section{The multiverse and experiments}

Apart from the Everett interpretation of quantum mechanics, many kinds of "multiverses" are now discussed in cosmology. One can cite for example the infinite space that would arise when $k=0$ or $k=-1$ (if the topology is simple) and contains infinitely many Hubble volumes, the interior of charged or rotating black holes if one takes seriously analytically extended Penrose-Carter diagrams, the inflationary bubbles possibly filled by different laws of physics, etc. Different laws of physics can in particular be obtained by compactification schemes and generalized magnetic fluxes in string theory or, more generically, by any kind of theory that has several local minima or vacua. We will, in the following of this section focus on such models. In each of the previously given examples, the word "universe" has a different meaning depending on the case.\\

Are those multiverses testable ? The answer is : in principle, yes. There are at least two different arguments in this direction. The first one is to emphasize (see, {\it e.g.}, \cite{carr} in particular arguments by M. Tegmark) that the multiverse is not a hypothesis but a consequence. It would indeed be an artificial and {\it ad hoc} hypothesis. But it actually appears as an output of theories that can be tested locally and that are built for completely other purposes, for example particle physics or quantum gravity. It is not an input. If those theories were to be falsified, all their predictions would disappear, including the multiverse. The other way round, if those theories were to be part of our main paradigm it would be inconsistant to discard  the multiverse they produce just because one does not like the idea. So, basically, the point is that we should not test the multiverse, but rather the theory that predicts it. And this is certainly in principle possible. Exactly as one uses general relativity (GR) to describe the internal structure of black holes even if this specific prediction cannot be tested, just because we are confident enough in GR to trust this.

The second argument is more closely related with the multiverse in itself. There is --in those approaches-- a large (possibly infinite) number of universes. And we see only a single one. Can something be said about the whole ensemble ? Obviously yes ! Let us imagine than instead of billions of particle collisions we were to see only one at the LHC. Would we have discovered the Higgs ? Of course not. But would be have been able to disproof a crazy model of particle physic? Most probably yes. A single sample contains less information than the full set but it does contain {\it some} information. It makes a statistical test possible, as usual in physics (either for quantum of classical reasons, all confirmations or exclusions are anyway statistical).To make precise predictions in the multiverse and compare them with our own universe, it is necessary but not sufficient to know the shape of the landscape. It is also necessary to account for a possible bias in the sample we observe (this is also very common in science). This precaution might be referred to as the anthropic principle. Let us make it very clear that there is absolutely no link at all between this principle (which is not actually a principle but rather a reminder) and intelligent design, anthropocentrism, theology or teleology. Just the other way round, the point is only to take care of the fact that as our planet is obviously not at all representative of our full universe, our universe might not be representative of the full multiverse. There is nothing mysterious or mystic here, this is just a bias that should be accounted for: we live in a place which is favorable to complexity, just because we are ourselves complex structure, of course we do not live in empty space or at the center of a neutron star. Similarly, our universe might be more complexity-friendly than most others and when trying to test the compatibility of a given theory with our observations we have to take care of that.

Let us assume that a model predicts a billion of universes, all of them but one being empty. Let us take a random sample. If this sample is full of stars, one can conclude that the theory is excluded at a high confidence level. But if we now consider a random sample selected from within by an observer and take into account the selection bias due to the fact that we are ourselves not made of vacuum, the model will not anymore be disfavored.\\

So, basically, when comparing the structure of the landscape (of solutions of the considered theory) with the sample universe we live in, the important quantity is the product  of the probability of a given solution $S_i$, $P(S_i)$, by the number of observer $N(S_i)$ in this solution. If one considers a random human being wondering where he lives, there are more chances that he is Indian rather than  Australian, even though Australia is larger than India, just because there are more Indian observers. This has to be accounted for when trying to reconstruct the landscapes of the Earth from local descriptions. It is well known that this formulation is obviously too vague and ill defined. We don't know how to define properly an observer in a general enough way. In addition, there are many severe limits associated with general problems in constructing a proper and unique measure. These are, however, mostly technical problems that can, in principle, be solved. Then, the compatibility between our world and this observer-weighted probability distribution over the landscape can be evaluated and the model excluded or confirmed at a given confidence level.\\

This view is not sterile. Even if one takes a narrow-minded non-ontological definition of physics (assumed to be only useful in making local predictions), this has consequences. As an example, one can consider the status of the inflationary model, in view of the Planck mission results. Depending on whether one takes into account the multiverse --that is a generic {\it prediction} of eternal inflation-- or ignore it, the compatibility of the model with data might change completely (see, {\it e.g.}, \cite{paul1}). Actually, our universe might even be exponentially disfavored in an inflationary multiverse \cite{paul2}, which is not at all the case in a single universe. 

\section{Basic idea for testing the Everett hypothesis}

Following \cite{page}, we would like to argue that the Everett interpretation of quantum mechanics can be tested in the very same way.\\

The question of testing the Everett vision has been raised long ago. Three possible "tests" that we won't describe in details here (see \cite{clive}) have been considered. The first one is related with linearity. In principle, because of linearity, it is possible to detect the presence of other nearby worlds, through the existence of interference effects. This has been pushed forward in \cite{deutsch} where an observer splits into two copies making different observations. They remember having made observations but not the results and can therefore be rejoined coherently into a single copy. This is very clever but nearly impossible to be made experimentally. 

The second one is that many-worlds actually requires that gravity be quantized, in contrast to other interpretations which are silent about the role of gravity. The reason is that if gravity was to remain non-quantized, all the universes that the Everett interpretation predicts should be easily detectable by their gravitational presence --they would all share the same background metric with our co-existing quantum worlds. Of course, gravity could be quantized and, still, the Everett interpretation could be wrong.

The third one, related with the first one and still using linearity (see \cite{deutsch2}) is based on reversible quantum computers. It requires currently beyond of reach artificial intelligence and reversible nanoelectronics.\\

What we want to emphasize here, as D. Page somehow did more than a decade ago, is that in the framework of quantum cosmology, testing the Everett interpretation of quantum mechanics is {\it not} different from testing any other multiverse proposal.\\

In a single universe interpretation of quantum mechanics, only the relative probability for different universes matters. There is only one universe, so only the probabilities between different outcomes are important. If the probability for a x-universe is much higher than the probability for a y-universe, we should be in the x-universe, whatever the number of observers it generates. However, in the Everett vision, all possible universes do actually exist. So, if the number of observers are different, say $N_x$ and $N_y$ respectively in the x-universe and the y-universe, the relevant probabilities are now the ones weighted by the number of observers. If the ratio $N_y/N_x$ is higher that $P(x)/P(y)$, the prediction is now the opposite. We should be in the y-universe. The point is that the situation is basically the same as in any multiverse situation. If there in only one World we should compute probability for this World, if there are many worlds the observer-weighted probability is the correct distribution. In principle, observations can select which one is true.\\

The methodology is very close to now standard approaches to testing the multiverse. Instead of just testing a model the point here is to test two interpretations of quantum mechanics. As long as the two interpretations lead to different final probability distributions, which is the case, the proposal is testable.

\section{Digression on what science is}

It is often argued in physics articles that "science is what Popper says: something that can be falsified, and we should not step away from that". First, we would like to underline that the multiverse is not outside this definition. It is {\it not} a theory, it is part of the predictions of theories that can in principle be falsified. As we explained before, many researches are devoted to the internal structure of black holes in GR. This is considered as usual science whereas nobody can go there and come back telling us if that is true. Once a model, in this exemple GR, is well enough tested to be part of the dominant paradigm, it is perfectly legitimate to use it even where it cannot be tested. It has hopefully never been necessary to check {\it all} the predictions of a theory to consider it as reliable and to use it.\\

Beyond that, we believe that we should not take Popper (or, more precisely this caricatural simplification of Popper's claims) too seriously. First, because it is only one epistemology among many others, say, for example (considering only the the XX$^{th}$ century), Carnap, Bachelard, Canguilhem, Cavaill\`es, Bourdieu, Feyerabend, Kuhn, Koyr\'e, Lakatos, Quine, Latour, Hacking, Merton, etc. All those epistemologies lead to consistent views on science that are completely different from the one of Popper whose popularity might just be due to its simplicity. But every scientist knows that falsifiability does not work in practice : all theories have troubles and depending on the situation we might assume invisible objects (dark matter for exemple) or a more limited domain of application of the theory to save it anyway. No serious theory has ever been refuted by a single experiment. Not to mention that most new theories were initially believed to be impossible to test/falsify (remember Auguste Comte believing, just before the discovery of spectroscopy, that discussing the composition of stars was not science as no one will ever be able to go there and analyse them). Popper's epistemology comes together with (or is deeply rooted in) a highly contestable firm belief that science will solve and has already solved most problems of the humanity ("today, misery has essentially disappeared from occidental Europe; excessive social disparity do not exist anymore", said he before criticizing "intellectuals" that do not agree with this obvious remark !).\\

Anyway, even if Popper was right in describing what science was --and we don't believe so--, say in the 1950's, why should we take this as a correct definition forever ? In essence, science is a dynamical process. Playing with the rules if obviously part of the game. Hopefully, the rules are not the same now as what they were for Galileo. Science is about deconstruction: nothing can be taken for granted, everything can --in principle-- be revised and changed. If one wants certainty and immuable rules, one should better consider theology. Of course, we all expect science to make predictions and to describe the World(s) more precisely. Everything is for sure not allowed. But one cannot know what are the correct revisions before exploring them and considering the consequences. So, even if the multiverse was --and we don't think it is the case-- to impose a change of the rules of the game, we think it would be more dangerous to {\it a priori} forbid the idea than to seriously study the new paradigm it might generate. All fields get redefined from within. This is obviously the case in art where contemporary musical composition, poetical writing and pictorial creation do not obey the aesthetics as it was defined initially by Baumgarten. Why wouldn't this evolution of the rules be allowed for science which is, probably more than any other, the cognitive field where changes are essential ?

\section{Specific possible tests of the Everett interpretation}

In \cite{page}, the particular example of the Hartle-Hawking no-boundary proposal was considered as a consistent proposal for a quantum state of the Universe. The Hartle-Hawking proposal basically does away with the initial three geometry, {\it i.e.} only includes four dimensional geometries that match onto the final three geometry. The path integral is interpreted as giving the probability of a universe with certain properties ({\it i.e.} those of the boundary three geometry) being created from essentially nothing. In a $S^3$ minisuperspace approach with a single $\phi^2$ potential  scalar field, the no-boundary proposal instanton leads to a set of FLRW universes with various number of inflationnary e-folds and therefore various properties,  with an amplitude given by:
$$
A\propto e^{D^2},
$$
$D$ being the linear spatial size of the Euclidean 4-dimensionnal hemisphere where the solution nucleates \cite{hawk}. Actually, $D\propto 1/\Phi_0$, where $\Phi_0$ is the initial value of the massive scalar field, and $N\propto exp(\Phi_0)$, where $N$ is the number of e-folds of inflation. Taking the volume of the resulting universe as a rough mesure proportional to the number of observers, one can compare the probability for our Universe being what it is in a single universe history on the one hand and, on the other hand, with the observer-weighted probability that should be chosen in the Everett many-worlds view where all universes do actually exist. The result strongly favors the Everett view. Obviously the conclusion should not be taken too seriously. Firstly, because it could very well be that the Hartle-Hawking proposal is not correct. Secondly, because even in this framework, the preliminary result from Page relies on many controversial assumptions. But this shows that testing the Everett proposal is possible.\\

This approach can be generalized to basically any model of quantum cosmology. The most delicate point is to define the measure for the number of observers for each outcome. This is quite straightforward in the case of closed universes (that are anyway often better defined in quantum cosmology) but very difficult in general. In the framework of Loop Quantum Cosmology \cite{ashtekar}, for example, the situation is more complicated. All quantum states can be shown to undergo a bounce in the sense that the expectation value of the volume operator has a non-zero lower bound in any state. Basically, the Wheel-DeWitt equation is replaced by a difference equation : $\partial_\varphi^2 \Psi (\nu, \varphi)=
 \frac{3\pi G}{4\lambda^2} \nu \left[\, (\nu+2\lambda)
\Psi(\nu+4\lambda) - 2\nu \tilde\Psi(\nu) + (\nu -2\lambda) \Psi(\nu-4\lambda) \right]$ where $\Psi$ is the wave function of the Universe and $ \lambda$ is the square root of the minimum area gap. The situation is tricky because at the semi-classical order, the duration of inflation is here determined by the {\it phase} of the inflaton field in the remote past, that is in the classical contracting branch. Although the most probable value can be determined \cite{linda}, this remains a random classical process. However, all quantum trajectories are still possible around the semi-classical approximation of highly peaked states. One has in principle all the required ingredients to calculate the weighted and unweighted probabilities for different universes and compare with ours.\\

Clearly, the procedure is still mostly out of reach just because we don't yet have an established theory of quantum cosmology (not to mention that some physicist even question the fact that quantum mechanics should apply to the Universe as whole \cite{ellis}) and it would be difficult to establish this theory if the interpretation of quantum mechanics if not fixed. The most promising avenue would be to have empirical evidence for a given quantum gravity proposal ({\it e.g.} by measuring Lorentz invariance violations, by detecting quantum gravity effects in black holes, or by any other means) and then apply this otherwise confirmed model to the Universe.\\

This proposal should be taken with care. It could probably be pushed {\it ad absurdum}. There is usually a non-vanishing quantum probability for basically everything. For example, there is a non-vanishing quantum probability for me to survive any kind of illness or aging process. In the Everett view, I will somehow be "immortal". Not for my friends and family, of course, seing me dying in most worlds. So I might argue that the simple fact that I am writing this article at an age which is not much above (and actually even smaller) that the mean life expectancy is already an argument disfavoring the model (the probability than I am less than 80 years-old if I am allowed to be eternal is very small). This would however forgets that the number of copies of myself decreases very fast as time goes on and this might counterbalance the naive argument.\\

The main point we wanted to make is that testing the Everett many-worlds is not fundamentally different from testing any multiverse model and is in principle possible.

\end{document}